\documentclass[twocolumn,prl,superscriptaddress,showpacs]{revtex4}

\usepackage{mathptmx}
\usepackage{helvet}
\usepackage{graphicx,fancyhdr}
\usepackage{ocg,dashbox,color}%
\usepackage{bm}
\usepackage{sidecap}%
\begin{document}

\title{Nanoscale layering of antiferromagnetic and superconducting phases in Rb$_{\mathbf{2}}$Fe$_{\mathbf{4}}$Se$_{\mathbf{5}}$}

\author{A.~Charnukha}
\affiliation{Max-Planck-Institut f\"ur Festk\"orperforschung, Heisenbergstrasse 1, D-70569 Stuttgart, Germany}
\author{A.~Cvitkovic}
\affiliation{Neaspec GmbH, D-82152 Martinsried (Munich), Germany}
\author{T.~Prokscha}
\affiliation{Laboratory for Muon Spin Spectroscopy, Paul Scherrer Institute (PSI), CH-5232 Villigen PSI, Switzerland}
\author{D.~Pr\"opper}
\affiliation{Max-Planck-Institut f\"ur Festk\"orperforschung, Heisenbergstrasse 1, D-70569 Stuttgart, Germany}
\author{N.~Ocelic}
\affiliation{Neaspec GmbH, D-82152 Martinsried (Munich), Germany}
\author{A.~Suter}
\author{Z.~Salman}
\author{E.~Morenzoni}
\affiliation{Laboratory for Muon Spin Spectroscopy, Paul Scherrer Institute (PSI), CH-5232 Villigen PSI, Switzerland}
\author{J.~Deisenhofer}
\affiliation{Experimental Physics V, Center for Electronic Correlations and Magnetism, Institute of Physics, University of Augsburg, D-86159 Augsburg, Germany}
\author{V.~Tsurkan}
\affiliation{Experimental Physics V, Center for Electronic Correlations and Magnetism, Institute of Physics, University of Augsburg, D-86159 Augsburg, Germany}
\affiliation{Institute of Applied Physics, Academy of Sciences of Moldova, MD-2028 Chisinau, R. Moldova}
\author{A.~Loidl}
\affiliation{Experimental Physics V, Center for Electronic Correlations and Magnetism, Institute of Physics, University of Augsburg, D-86159 Augsburg, Germany}
\author{B.~Keimer}
\author{A.~V.~Boris}
\affiliation{Max-Planck-Institut f\"ur Festk\"orperforschung, Heisenbergstrasse 1, D-70569 Stuttgart, Germany}

\begin{abstract}
We studied phase separation in a single-crystalline antiferromagnetic superconductor $\textrm{Rb}_2\textrm{Fe}_4\textrm{Se}_5$ (RFS) using a combination of scattering-type scanning near-field optical microscopy (s-SNOM) and low-energy muon spin rotation (LE-$\mu$SR). We demonstrate that the antiferromagnetic and superconducting phases segregate into nanometer-thick layers perpendicular to the iron-selenide planes, while the characteristic in-plane size of the metallic domains reaches 10~$\mu$m. By means of LE-$\mu$SR we further show that in a 40-nm thick surface layer the ordered antiferromagnetic moment is drastically reduced, while the volume fraction of the paramagnetic phase is significantly enhanced over its bulk value. Self-organization into a quasiregular heterostructure indicates an intimate connection between the modulated superconducting and antiferromagnetic phases.

\end{abstract}


\maketitle
The recent discovery of intercalated iron-selenide superconductors~\cite{PhysRevB.82.180520,PhysRevB.83.212502,APL10.10631.3549702,PhysRevB.83.060512,NatMatZhangFengARPESnohole2011} has stirred up the condensed-matter community accustomed to the proximity of the superconducting and magnetic phases in various cuprate and pnictide superconductors. Never before has a superconducting state with a transition temperature as high as 30~K been found to coexist with such an exceptionally strong antiferromagnetism with N\'eel temperatures up to 550~K as in this new family of iron-selenide materials. The very large magnetic moment of $3.3\ \mu_{\mathrm{B}}$ on the iron sites~\cite{Bao_KFS_2011}, however, renders a microscopically homogeneous coexistence of superconductivity and magnetism unlikely. Indeed, significant experimental evidence suggests that the superconducting and antiferromagnetic phases are spatially separated~\cite{PhysRevB.84.060511,PhysRevX.1.021020,SciRep_Wang2012,2011arXiv1108.5698C,PhysRevB.84.180508,KFS_MBE_thinfilm_NatPhys_2011,PhysRevB.84.094504}.
\begin{figure*}[!th]
\includegraphics[width=\textwidth]{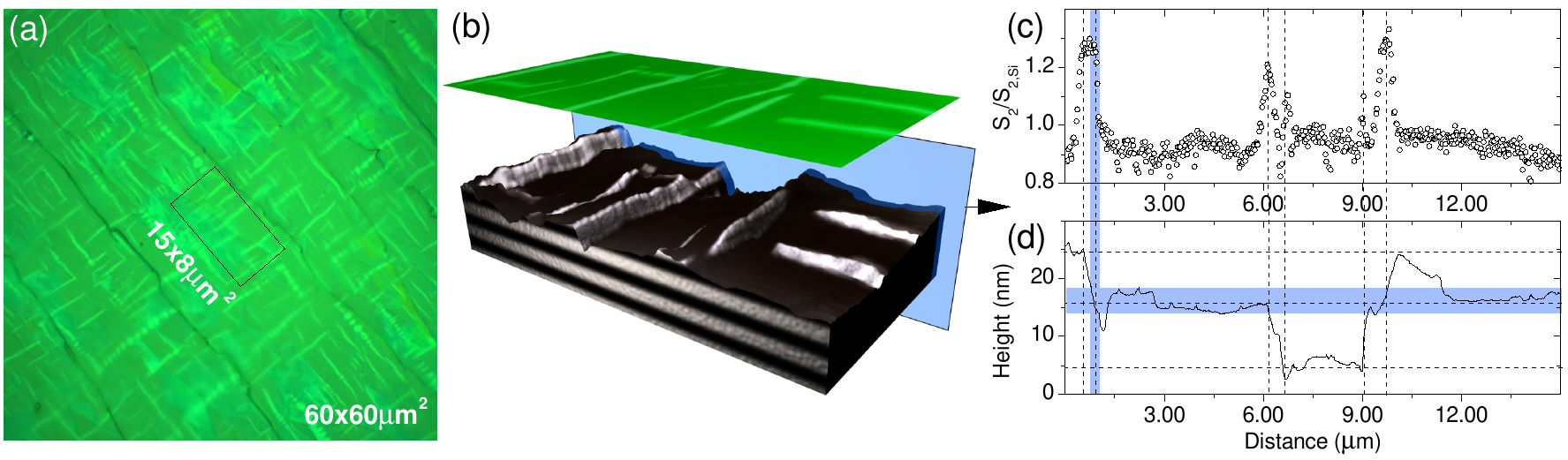}
\caption{\label{fig:neasom} (a)~Microscope image of a $60\times60\ \mu\textrm{m}^2$ surface patch of a freshly-cleaved superconducting RFS single crystal. Typical rectangular $15\times8\ \mu\textrm{m}^2$ area studied via near-field microscopy. (b)~Superposition of the topography of a $15\times8\ \mu\textrm{m}^2$ rectangular area (terrain) and the optical signal (brightness) normalized to that of silicon. Glossy areas indicate high silicon-RFS contrast and thus metallicity, while the matt areas are insulating. This combined response is broken down in~(c) and~(d) for the cross-section defined by the blue semi-transparent plane. (c)~Optical contrast $S_2/S_\mathrm{2,Si}$ of the second harmonic $S_2$ of the near-field signal obtained at the $10.7\ \mu\textrm{m}$ emission wavelength ($116\ \textrm{meV}$ photon energy) of a $\textrm{CO}_2$ laser. Peaks in the contrast indicate metallic response (see text). (d)~Displacement of the AFM tip while scanning along a $15\ \mu\textrm{m}$ line obtained simultaneously with~\textbf{c}. Dashed lines and blue shaded areas in~(c) and~(d) show the correlation between the metallic response and changes in the topography.}
\end{figure*}
\par In the prototypical iron arsenide superconductors $\textrm{Ba(K)}\textrm{Fe(Co)}_2\textrm{As}_2$ both the phase separation~\cite{PhysRevLett.102.117006,PhysRevB.79.224503} and coexistence~\cite{PhysRevLett.105.057001} of antiferromagnetism and superconductivity have been shown to occur in certain regions of the phase diagram. At the same time, the structurally similar intercalated iron-selenide compounds $\textrm{(K,Rb,Cs)}_{0.8}\textrm{Fe}_{1.6}\textrm{Se}_2$ have defied all the efforts to synthesize a bulk single-phase material of this family. Absence of such electronically homogeneous superconducting single crystals and a strong correlation between the superconducting and antiferromagnetic phases~\cite{PhysRevLett.107.137003,2011arXiv1112.3822K} necessitate a detailed research into the nature of their coexistence. The volume fraction of the magnetic phase has been estimated to $88\%$ in recent M\"ossbauer~\cite{PhysRevB.84.180508} and bulk $\mu$SR~\cite{2011arXiv1111.5142S} studies. The shape of the phase domains, on the contrary, has seen much conflicting evidence with indications ranging from needlelike rather regular stripes~\cite{SciRep_Wang2012} to insulating islands on a superconducting surface~\cite{PhysRevX.1.021020} to nanoseparated vacancy-disordered presumably metallic sheets in the bulk with an unknown in-plane form factor~\cite{PhysRevB.83.140505}. In a recent STM study of [110] $\textrm{K}_x\textrm{Fe}_{2-y}\textrm{Se}_2$~(KFS) thin films the superconducting phase was assigned to stoichiometric $\textrm{KFe}_2\textrm{Se}_2$ without iron vacancies~\cite{KFS_MBE_thinfilm_NatPhys_2011}. However, consistent theoretical description of the recent inelastic neutron scattering and angle-resolved photoemission measurements based on this assumption requires significantly different levels of the chemical potential in the bulk and at the surface (equivalent to a disparity of $\sim0.08$~electrons/Fe in the electron doping)~\cite{2011arXiv1112.1636F,PhysRevB.83.100515}.
\par Infrared spectroscopy is well-suited to clearly distinguish between the antiferromagnetic semiconducting and paramagnetic metallic phases of RFS due to a large contrast in its complex dielectric function in this spectral range~\cite{2011arXiv1108.5698C}. To determine the geometry of the domains and unambiguously assign them to the semiconducting and metallic phases, a submicrometer-resolution technique must be used. In this Letter we employed apertureless, scattering-type scanning near-field optical microscopy (s-SNOM) in the infrared~\cite{Hillenbrand_sSNOM2002,Keilmann_Hillenbrand_book,NatMat_10_352_2011}, which enables determination of the material's complex dielectric function with an unsurpassed in-plane resolution of ca $20\ \textrm{nm}$ and a typical nanometer topographic sensitivity of an atomic-force microscope (AFM). We complemented this spatially-resolved technique with low-energy muon spin rotation (LE-$\mu$SR)~\cite{Prokscha2008317,Boris_Science_Superlattice_2011} measurements on the same RFS single crystals to quantify the fraction of the magnetically-ordered phase in the bulk and trace its modification towards the sample surface.
\par The commercial s-SNOM near-field microscope uses an illuminated AFM probing tip to scan the sample surface and pseudo-heterodyne interferometric detection~\cite{Ocelic_Pseudohet_detection} to extract the near-field amplitude and phase from the light scattered back from the tip. Standard platinum-coated AFM tips (NanoWorld ARROW-NCPt with a $25\ \textrm{nm}$ radius) were used as antennas for nanofocusing of $\textrm{CO}_2$ laser radiation at $10.7\ \mu\textrm{m}$ ($116\ \textrm{meV}$) wavelength (photon energy). The measurements were carried out on optimally-doped superconducting RFS single crystals (batch BR26 in Ref.~\onlinecite{PhysRevB.84.144520}, $T_{\textrm{c}}\approx32\ \textrm{K}$), cleaved prior to every scan. The sample surface was first characterized with a polarizing microscope. Figure~\ref{fig:neasom}(a) shows a $60\times60\ \mu\textrm{m}^2$ patch of the sample surface. A network of bright stripes is always observed on a freshly cleaved surface and does not depend on the polarization of the probing light. The stripes always occur at $45^\circ$ with respect to the in-plane crystallographic axes. The topography of a representative $15\times8\ \mu\textrm{m}^2$ surface patch studied with an s-SNOM microscope is shown in Fig.~\ref{fig:neasom}(b) as a three-dimensional terrain image. One can clearly correlate the features in the AFM map with the bright stripes in the polarizing microscope image. This implies that the surface chemistry of this compound leads to inherent surface termination with mesoscopic terracing upon cleaving, typically $10-30\ \textrm{nm}$ high. Figures~\ref{fig:neasom}(c) and~\ref{fig:neasom}(d) show the 2nd-harmonic near-field optical contrast (OC), obtained by normalizing the signal from the RFS surface to that that from a reference silicon surface ($S_2/S_\mathrm{2,Si}$), and the topography profile of RFS for the cross-section indicated with a blue translucent plane in Fig.~\ref{fig:neasom}(b). The amplitude and phase of both the topography and the optical signal are obtained simultaneously during a scan. Every peak in the OC maps signals a metallic optical response. The absolute values of the complex dielectric function were obtained within the extended finite-dipole model~\cite{Cvitkovic_07} using OC maps. The dielectric response of the semiconducting phase [dark regions in Fig.~\ref{fig:neasom}(b) and low OC in Fig.~\ref{fig:neasom}(c)] obtained in such a treatment of the experimental data ($\varepsilon_1\approx10,\ \varepsilon_2\approx0$) is fully consistent with that of the single-phase semiconducting RFS crystals~\cite{2011arXiv1108.5698C}. The bright regions of the sample surface display negative values of $\varepsilon_1$, which provides solid evidence for their metallic character.
\begin{figure*}[!th]
\includegraphics[width=\textwidth]{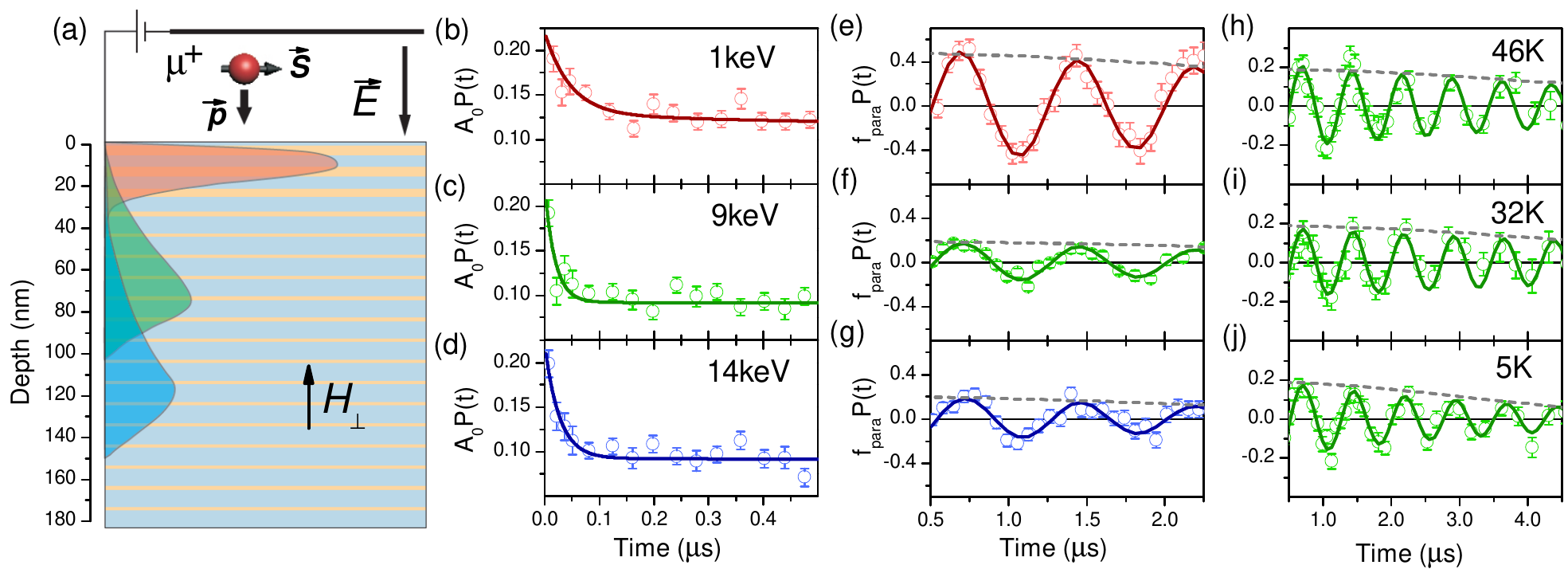}
\caption{\label{fig:musr} (a)~Schematic of the experiment with low-energy muons. The sample is depicted as a layered structure according to the discussion in the text. (b)-(d)~Time dependence of $\mu^+$ spin polarization $A_0P(t)$ in a zero magnetic field at $5\ \textrm{K}$ after implantation in RFS single crystals at various depths determined by the muon stopping profiles of respective colors in~(a). (e)-(g)~Same for normalized $\mu^+$ spin polarization $f_{\mathrm{para}}P(t)$ in a transverse magnetic field $H_\perp=100\ \textrm{G}$, where the normalized asymmetry $f_{\mathrm{para}}$ is the paramagnetic volume fraction. (h)-(j)~Long-time dependence of normalized $\mu^+$ spin polarization $f_{\mathrm{para}}P(t)$ in a transverse magnetic field $H_\perp=100\ \textrm{G}$ for the green muon stopping profile in~(a) in the normal state at $46\ \textrm{K}$~(h) and~$32\ \textrm{K}$~(i) and superconducting state at $5\ \textrm{K}$~(\textbf{j}). Gray dashed lines in~(e)-(j) show the slow relaxation envelope of the $\mu^+$ spin polarization; solid lines are fits to the data.}
\end{figure*}
\par By correlating the OC peak positions with the topography maps one can deduce that the location of the metallic regions on the sample surface is bound to the slopes in the terrain, as indicated in Figs.~\ref{fig:neasom}(c) and~\ref{fig:neasom}d with dashed lines. Flat regions of the sample surface exhibit no metallic regions. Figure~\ref{fig:neasom}(d) shows that from the location of the OC peaks on the slopes one can identify well-defined sheets (horizontal dashed lines) approximately $10\ \textrm{nm}$ apart. Metallic response (OC peak) is recorded whenever one of these sheets is exposed from the sample bulk at the intersections of the dashed lines with the topography line. Such correlation exists in all studied cross-sections of the OC and height maps. The thickness of the metallic sheets can be estimated by projecting the full width at half maximum (FWHM) of the OC peaks onto the surface topography [blue shaded areas in Fig.~\ref{fig:neasom}(c) and~\ref{fig:neasom}(d)] and on average amounts to $5\ \textrm{nm}$. The analysis of the correlations between the OC peaks and the slopes in the sample topography carried out on multiple cross-sections of different near-field maps shows that the metallic sheets have an in-plane dimension as well, which can be roughly estimated to $\approx10\ \mu\textrm{m}$ in both directions. Therefore, phase separation in superconducting RFS single crystals occurs on the nanoscale in the out-of-plane and mesoscale in the in-plane direction. From the thickness and periodicity of the metallic sheets we can estimate the volume fraction of the metallic domains in the $30\ \textrm{nm}$ surface layer to approximately $50\%$, significantly larger than all previously reported bulk values. 
\par It must be noted that the metallic volume fraction obtained by means of s-SNOM imaging only considers the near surface layer within $\approx30\ \textrm{nm}$. Whether or not this fraction changes with the distance from the sample surface and what the magnetic and superconducting properties of the metallic and semiconducting phases are cannot be determined from these measurements. Therefore, a microscopic sensor of the local magnetic moment with an adjustable implantation depth must be invoked. Such characteristics are provided by the LE-$\mu$SR technique, which utilizes the predominantly spin-oriented positron decay of $\mu^+$ to detect the orientation of the muon spin. This orientation is influenced by the local magnetic field at the muon implantation site, as well as magnetic fields applied externally. The experiment is shown schematically in Fig.~\ref{fig:musr}(a). Based on the results of the near-field measurements the sample is sketched as an idealized periodic layered structure, the finite in-plane dimension of the paramagnetic domains and the irregularity of their periodicity having been disregarded at this point.
\par In the LE-$\mu$SR technique, in contrast to its conventional counterpart, the incident energetic ($\sim\textrm{MeV}$) muon beam is first moderated to about $15\ \textrm{eV}$ in a condensed layer of solid $\textrm{N}_2$ or $\textrm{Ar}$ and then accelerated by a controlled electric field to achieve an adjustable implantation depth according to the muon stopping profile in a given material, calculated using the Monte Carlo algorithm TRIM.SP~\cite{Morenzoni_stopping_profile_2002}. To study the dependence of the phase separation in RFS single crystals on the distance from the sample surface, we chose three muon stopping profiles shown as a red (achieved by accelerating the moderated muon beam to the energy of $1\ \textrm{keV}$), green ($9\ \textrm{keV}$), and blue ($14\ \textrm{keV}$) shaded profile in Fig.~\ref{fig:musr}(a). The resulting time dependence of $\mu^+$ spin polarization $A_0P(t)$ at zero external magnetic field is shown in Figs.~\ref{fig:musr}(b)-(d), the colors correspond to those of the muon stopping profiles in Fig.~\ref{fig:musr}(a). As is clear from Fig.~\ref{fig:musr}(a), the shallowest muon stopping profile (red shaded area) is peaked at about $10\ \textrm{nm}$ and extends to about $30\ \textrm{nm}$ below the sample surface. It thus probes a region of the sample bulk comparable to that observed via s-SNOM. The relaxation of the muon polarization after implantation in the sample occurs at two different time scales: fast depolarization takes place during the first $150-250\ \textrm{ns}$ and is associated with the ordered antiferromagnetic phase, while a much slower evolution dominates thereafter and stems predominantly from the paramagnetic volume fraction. The fast depolarization rate is proportional to the width of the magnetic-field distribution at the muon stopping site in the ordered phase and thus to the antiferromagnetic moment. Figures~\ref{fig:musr}(c) and~\ref{fig:musr}(d) show that in the bulk [green and blue muon stopping profiles in Fig.~\ref{fig:musr}(a)] the fast component is approximately the same and rather narrow, whereas near the surface it broadens significantly [Fig.~\ref{fig:musr}(b)]. Using a two-component fit model for the zero-field $\mu$SR data, the fast depolarization rates in the bulk were found to agree within the error bars: $52(13)\ \mu\textrm{s}$ and $37(7)\ \mu\textrm{s}$ for the green and blue profiles, respectively, whereas it is significantly reduced to $21(4)\ \mu\textrm{s}$ close to the surface (red profile). Under the assumption that the muon stopping sites in the unit cell are the same for all three profiles and that the depolarization rate should be the same in the bulk, the antiferromagnetic moment is reduced to only $50\%$ of its bulk value in the $30\ \textrm{nm}$ surface layer. 
\par The muon decay asymmetry in Figs.~\ref{fig:musr}(b)-(d) indicates that the paramagnetic volume fraction (slow component) is enhanced closer to the surface. To avoid the contaminating $\mu$SR signal originating from muon decays in the Ni sample holder, we carried out transverse-field measurements in the same configuration and at the same temperature of $5\ \textrm{K}$ as in Figs.~\ref{fig:musr}(b)-(d). In this case the oscillating component of the muon decay asymmetry comes only from the paramagnetic domains in the sample and can be normalized accordingly to become $f_{\mathrm{para}}P(t)$, where $f_{\mathrm{para}}$ is the paramagnetic volume fraction, given by the value of the slowly depolarizing envelope of the muon spin precession (gray dashed lines in Fig.~\ref{fig:musr}) at zero time. From Figs.~\ref{fig:musr}(f) and~\ref{fig:musr}(g) one immediately infers that well in the sample bulk the paramagnetic phase constitutes about $20\%$ of the sample volume. It thus characterizes the phase separation in the bulk of the superconducting RFS single crystals as probed by other techniques, such as M\"ossbauer spectroscopy and conventional $\mu$SR. On the other hand, Fig.~\ref{fig:musr}(e) shows that in the $30\ \textrm{nm}$ surface layer [red stopping profile in Fig.~\ref{fig:musr}(a)] the paramagnetic volume fraction strongly increases to $\approx50\%$. Such a high value most likely occurs due to a reduced constraining potential between antiferromagnetic sheets with a significantly smaller ordered moment close to the sample surface, which leads to an expansion of the metallic regions. These depth-dependent physical properties could explain the difference in the chemical potential required for a consistent interpretation of the data obtained with bulk (inelastic neutron scattering) and surface-sensitive (ARPES) probes.
\par To characterize the superconducting phase of RFS with $\mu$SR we studied the temperature dependence of the oscillating component of the muon asymmetry in a vortex state generated by the transverse external magnetic field, perpendicular to the metallic sheets. The oscillations are damped by the inhomogeneity of the magnetic field in the vortices, which, in turn, is proportional to the condensate density $n_{\mathrm{s}}\propto1/\lambda_{\mathrm{ab}}^2$. The measurements were carried out in a transverse magnetic field $H_\perp=100\ \textrm{G}$ for the muon stopping profile peaked at $80\ \textrm{nm}$ below the sample surface [green shaded area in Fig.~\ref{fig:musr}(a)]. The time dependence of the normalized $\mu^+$ spin polarization $f_{\mathrm{para}}P(t)$ in the normal state at $46$ and $32\ \textrm{K}$ and in the superconducting state at $5\ \textrm{K}$ is shown in Figs.~\ref{fig:musr}(h)-(j), respectively. It is evident from the data that the damping in the normal state is approximately constant, while in the superconducting state it is noticeably faster, which indicates the presence of a superconducting condensate. By subtracting the normal-state damping from that in the superconducting state we can estimate the London penetration depth $\lambda_{\mathrm{||}}\approx550\ \textrm{nm}$. Assuming that the phase separation in the bulk has the same layered structure as detected near the surface, this value must be revisited in an appropriate model. For a stack of superconducting layers of thickness $d_{\mathrm{sc}}$ separated by insulating layers of thickness $d_{\mathrm{ins}}$ the inhomogeneity of the magnetic field in a vortex (which is then a stack of two-dimensional vortex pancakes) is described by the Lawrence-Doniach in-plane penetration depth $\lambda_{\mathrm{||}}$, related to the bulk London penetration depth $\lambda_{\mathrm{ab}}$ via $\lambda_{\mathrm{||}}=\lambda_{\mathrm{ab}}(d_{\mathrm{sc}}+d_{\mathrm{ins}})^{1/2}/d_{\mathrm{sc}}^{1/2}$ (see Ref.~\onlinecite{PhysRevB.43.7837}). This reduction reflects the fact that $\lambda_{\mathrm{||}}$ is related to the {\it average} superconducting condensate density $\langle n_{\mathrm{s}}\rangle=n_{\mathrm{s}}d_{\mathrm{sc}}/(d_{\mathrm{sc}}+d_{\mathrm{ins}})=n_{\mathrm{s}}f_{\mathrm{para}}$. Taking these considerations into account and using the bulk paramagnetic volume fraction $f_{\mathrm{para}}=0.2$ obtained in our LE-$\mu$SR measurements one can estimate the intrinsic bulk in-plane London penetration depth of the superconducting phase to be $\lambda_{\mathrm{ab}}\approx250\ \textrm{nm}$. This value would increase towards $\lambda_{\mathrm{||}}$ due to the finite in-plane dimension of the superconducting layers and their disordered stacking since the superconducting phase inclusions would then contribute independently to the $\mu$SR depolarization. At the same time, our procedure slightly overestimates the value of $\lambda_{\mathrm{||}}$ due to the widening of the vortex field distribution close to the surface~\cite{PhysRevLett.83.3932}.
\par Similar effective-medium approximations (EMA) must be used for an adequate analysis of the results obtained with other experimental techniques. A recent study of the optical conductivity of the same superconducting single crystals~\cite{2011arXiv1108.5698C} emphasized the importance of EMA but could not make a concrete estimate due to the unknown details of the phase separation in this compound. It reported the total plasma frequency of the itinerant charge carriers of about $100\ \textrm{meV}$. Using the same dimensions for the metallic paramagnetic and semiconducting antiferromagnetic layers as in the interpretation of the LE-$\mu$SR results one can extract the bulk superconducting optical response from the conductivity data reported in Ref.~\onlinecite{2011arXiv1108.5698C}. A simple fit of the experimental data as the optical response of a perfect superlattice gives a significantly larger value of the total plasma frequency $\omega_{\mathrm{pl}}^{\mathrm{tot}}\approx300\ \textrm{meV}$ and brings the estimated value of the London penetration depth $\lambda_{\mathrm{ab}}^{\mathrm{opt}}\approx2\ \mu\textrm{m}$ closer to that obtained with LE-$\mu$SR but still 8 times larger. Accounting for the finite in-plane dimension of the paramagnetic domains is expected to lead to a better agreement between the two techniques.
\par The origin of the phase separation observed in this work can lie either in a chemical stratification into e.g. iron-vacancy ordered (antiferromagnetic) and disordered (paramagnetic) phases or in a purely electronic segregation on a homogeneous crystalline background. Self-organization of a chemically homogeneous structure into (quasi)periodically segregated phases is not unprecedented --- a similar phenomenon has been observed in copper-based superconductors, where antiferromagnetic stripes of copper spins were found to be spatially separated by periodic domain walls close to a particular hole doping level of $1/8$~\cite{Tranquada_Uchida_stripe_domains_1995}. Be it of chemical or electronic nature, the phase separation in superconducting RFS single crystals reported here represents an interesting case of a naturally-occurring quasi-heterostructure.
\par To conclude, by combining the unique optical imaging capabilities and nanoscale resolution of the s-SNOM near-field microscope with bulk sensitivity at variable depth of LE-$\mu$SR we determined the geometry and magnitude of the phase separation in RFS superconducting single crystals. The paramagnetic domains were found to have a shape of thin metallic sheets parallel to the iron-selenide plane of the crystal with a characteristic size of only several nanometers out of plane but up to 10~$\mu$m in plane. By means of LE-$\mu$SR we further show that the antiferromagnetic semiconducting phase occupies $\approx80\%$ of the sample volume in the bulk and is strongly weakened near the surface. These results have important implications for the interpretation of bulk- and surface-sensitive measurements on $\textrm{Rb}_2\textrm{Fe}_4\textrm{Se}_5$, and for the understanding of the interplay between superconductivity and antiferromagnetism in this material.

\begin{thebibliography}{}

\bibitem{PhysRevB.82.180520}
J. Guo \textit{et~al.}, Phys. Rev. B \textbf{82}, 180520 (2010).

\bibitem{PhysRevB.83.212502}
J.~J. Ying \textit{et~al.}, Phys. Rev. B \textbf{83}, 212502 (2011).

\bibitem{APL10.10631.3549702}
Y. Mizuguchi \textit{et~al.}, Appl. Phys. Lett. \textbf{98}, 042511 (2011).

\bibitem{PhysRevB.83.060512}
A.~F. Wang \textit{et~al.}, Phys. Rev. B \textbf{83}, 060512 (2011).

\bibitem{NatMatZhangFengARPESnohole2011}
Y. Zhang \textit{et~al.}, Nature Mater. \textbf{10}, 273--277 (2011).

\bibitem{Bao_KFS_2011}
W. {Bao} \textit{et~al.}, Chinese~Phys.~Lett. \textbf{28}, 086104 (2011).

\bibitem{PhysRevB.84.060511}
A. {Ricci} \textit{et~al.}, Phys.~Rev.~B \textbf{84}, 060511 (2011).

\bibitem{PhysRevX.1.021020}
F. Chen \textit{et~al.}, Phys. Rev. X \textbf{1}, 021020 (2011).

\bibitem{SciRep_Wang2012}
R.~H. {Yuan} \textit{et~al.}, Sci.~Rep. \textbf{2}, 221 (2012).

\bibitem{2011arXiv1108.5698C}
A. {Charnukha} \textit{et~al.}, arXiv:1108.5698 (unpublished) (2011).

\bibitem{PhysRevB.84.180508}
V. Ksenofontov \textit{et~al.}, Phys. Rev. B \textbf{84}, 180508 (2011).

\bibitem{KFS_MBE_thinfilm_NatPhys_2011}
W. {Li} \textit{et~al.}, Nature~Phys. \textbf{8}, 126--130 (2012).

\bibitem{PhysRevB.84.094504}
M. Wang \textit{et~al.}, Phys. Rev. B \textbf{84}, 094504 (2011).

\bibitem{PhysRevLett.102.117006}
J.~T. Park \textit{et~al.}, Phys. Rev. Lett. \textbf{102}, 117006 (2009).

\bibitem{PhysRevB.79.224503}
D.~S. Inosov \textit{et~al.}, Phys. Rev. B \textbf{79}, 224503 (2009).

\bibitem{PhysRevLett.105.057001}
P. Marsik \textit{et~al.}, Phys. Rev. Lett. \textbf{105}, 057001 (2010).

\bibitem{PhysRevLett.107.137003}
F. Ye \textit{et~al.}, Phys. Rev. Lett. \textbf{107}, 137003 (2011).

\bibitem{2011arXiv1112.3822K}
V. {Ksenofontov} \textit{et~al.}, arXiv:1112.3822 (unpublished) (2011).

\bibitem{2011arXiv1111.5142S}
Z. {Shermadini} \textit{et~al.}, arXiv:1111.5142 (unpublished) (2011).

\bibitem{PhysRevB.83.140505}
Z. Wang \textit{et~al.}, Phys. Rev. B \textbf{83}, 140505 (2011).

\bibitem{2011arXiv1112.1636F}
G. {Friemel} \textit{et~al.}, arXiv:1112.1636 (unpublished) (2011).

\bibitem{PhysRevB.83.100515}
T.~A. Maier \textit{et~al.}, Phys.~Rev.~B \textbf{83}, 100515 (2011).

\bibitem{Hillenbrand_sSNOM2002}
R. Hillenbrand, T. Taubner and F. Keilmann, Nature \textbf{418}, 159--162
  (2002).

\bibitem{Keilmann_Hillenbrand_book}
F. Keilmann and R. Hillenbrand, \textit{Nano-Optics and Near-Field Optical Microscopy}
  (Artech House, 2008).

\bibitem{NatMat_10_352_2011}
F. Huth \textit{et~al.}, Nature~Mater. \textbf{10}, 352--356 (2011).

\bibitem{Prokscha2008317}
T. Prokscha \textit{et~al.}, Nucl. Instrum. Methods Phys. Res., Sect. A
  \textbf{595}, 317--331 (2008).

\bibitem{Boris_Science_Superlattice_2011}
A.~V. Boris \textit{et~al.}, Science \textbf{332}, 937--940 (2011).

\bibitem{Ocelic_Pseudohet_detection}
N. Ocelic, A. Huber and R. Hillenbrand, Appl.~Phys.~Lett. \textbf{89}, 101124
  (2006).

\bibitem{PhysRevB.84.144520}
V. Tsurkan \textit{et~al.}, Phys. Rev. B \textbf{84}, 144520 (2011).

\bibitem{Cvitkovic_07}
A. Cvitkovic, N. Ocelic and R. Hillenbrand, Opt. Express \textbf{15},
  8550--8565 (2007).

\bibitem{Morenzoni_stopping_profile_2002}
E. Morenzoni \textit{et~al.}, Nucl. Instrum. Methods Phys. Res. B \textbf{192},
  254--266 (2002).

\bibitem{PhysRevB.43.7837}
J.~R. Clem, Phys. Rev. B \textbf{43}, 7837--7846 (1991).

\bibitem{PhysRevLett.83.3932}
C. Niedermayer \textit{et~al.}, Phys. Rev. Lett. \textbf{83}, 3932 (1999).

\bibitem{Tranquada_Uchida_stripe_domains_1995}
J.~M. Tranquada \textit{et~al.}, Nature \textbf{375}, 561--563 (1995).

\end{thebibliography}

\end{document}